\documentclass[a4paper,10pt]{article}
\usepackage{epsfig}
\usepackage{amssymb}
\usepackage{graphicx}
\usepackage{graphicx,floatflt,amssymb} 
\usepackage{epsfig} 
\usepackage{amssymb,amsmath}

\def\circa#1{\,\raise.3ex\hbox{$#1$\kern-.75em\lower1ex\hbox{$\sim$}}\,}

\newcommand{\eV}{\,{\rm eV}}

\newcommand{\GeV}{\,{\rm GeV}}

\makeatletter
\def\art{\@ifnextchar[{\eart}{\oart}}
\def\eart[#1]#2#3#4#5#6{{\rm #2}, {\em #3 \rm #4} {\rm (#6) #5 ({\em #1})}}
\def\hepart[#1]#2{{\rm #2, \em#1}}
\newcommand{\oart}[5]{{\rm #1}, {\em #2 \rm #3} {\rm (#5) #4}}

\newcommand{\beq}{\begin{equation}}
\newcommand{\eeq}{\end{equation}}
\newcommand{\bea}{\begin{eqnarray}}
\newcommand{\eea}{\end{eqnarray}}
\newcommand{\ba}{\begin{array}}
\newcommand{\ea}{\end{array}}
\newcommand{\bi}{\begin{itemize}}
\newcommand{\ei}{\end{itemize}}
\newcommand{\bn}{\begin{enumerate}}
\newcommand{\en}{\end{enumerate}}
\newcommand{\bc}{\begin{center}}
\newcommand{\ec}{\end{center}}

\newcommand{\gsim}{\lower.7ex\hbox{$\;\stackrel{\textstyle>}{\sim}\;$}}
\newcommand{\lsim}{\lower.7ex\hbox{$\;\stackrel{\textstyle<}{\sim}\;$}}

\begin{document}
\tolerance=100000
\thispagestyle{empty}
\setcounter{page}{0}

\begin{flushright}

\texttt{LPT-Orsay/06-61}\\

\end{flushright}

\begin{center}
{\large \bf 
{Study of flavour dependencies in leptogenesis}}\\[0.3cm]
\vskip 10pt
{\bf F. X. Josse-Michaux,  A. Abada  
}
 
\vskip 5pt  
{\it Laboratoire de Physique Th\'eorique, UMR 8627,  
Universit\'e de Paris-Sud 11, B\^atiment 210,\\ 91405 Orsay Cedex,
France} \\

\vspace{0.5cm}
{\bf Abstract}
\end{center}
\begin{quote}
{\noindent
{We study the impact of flavours  on the efficiency factors  and give analytical and numerical results of the baryon asymmetry taking into account  the different charged lepton Yukawa contributions and the complete 
(diagonal and off-diagonal) $L$ to $B-L$ conversion  $A$ matrix. With this  treatment we update the lower bound on the lightest right-handed neutrino mass.}}
\end{quote}

\vspace{1cm}
\setcounter{page}{1}
\section{Introduction}
The minimal leptogenesis scenario \cite{FY} is based on the seesaw (type I) mechanism, consisting
of the Standard Model (SM) extended by 2 or 3 right-handed (RH) Majorana neutrinos with hierarchical masses. The 
lightest RH neutrino, produced by thermal scattering after inflation, decays through out-of-equilibrium processes that violate lepton number, C  and CP symmetries. These processes induce a dynamical production of a lepton asymmetry, which can be converted into a 
baryon asymmetry through 
$(B+L)$-violating  sphaleron  interactions.
In this context, several studies 
 \cite{leptogen,towards,pedes} investigated  this possible explanation of the baryon asymmetry of the Universe (BAU)
  and the different analyses have led to 
 constraints on neutrino physics from the requirement of a successfull leptogenesis. 
For instance,  a lower bound on  the reheating temperature 
$T_{\rm RH}$  (see e.g. \cite{towards,towardsbis,DI}) and an upper bound on
the absolute scale of light neutrino masses (for example in \cite{bound}) have been derived.
These studies have been performed using the so called ``one-flavour state" approximation. 
It is becoming well known  that the explanation of the BAU from a successfull thermal leptogenesis 
 has to be revisited when the mass of the decaying right-handed  neutrino that produces the lepton asymmetry  is  $M_{N_{1}} \lesssim 10^{12}$ GeV \cite{flav,zeno}. In this case, the Yukawa couplings of the charged leptons affect  the dynamics of the Boltzmann equations (BE) and one cannot use the usual one-state-dominance approximation (one flavour), but the ``flavoured" dynamical equations in the derivation of the BAU must be taken into account~ \cite{flav}-\cite{zeno}. 
 \section{Flavours in  leptogenesis}
We consider the SM Lagrangian with three heavy Majorana right-handed neutrinos $N_i$, $i=1,2,3$. The RH neutrinos couple to the left-handed (LH) ones through the complex  Yukawa coupling matrix, $\lambda$. The small neutrino masses generated by the seesaw (type I) mechanism are given by~:
\begin{eqnarray}
m_{\nu}=v^2\, U^{T}\,\lambda^{T}\,{\cal M }^{-1}\,\lambda\,U\ ,
\end{eqnarray}
where $U$ is the PMNS mixing matrix, $v$ is the vacuum expectation value (vev) of the Higgs field ($v\simeq 174$ GeV), and ${\cal M}$ is the $3\times 3$ diagonal Majorana mass matrix.  
We assume a hierarchical spectrum for the right handed neutrino, $M_{N_{3}}\gg M_{N_{2}} \gg M_{N_{1}}$, and  consider that  the lepton asymmetry is produced by the decay of the lightest RH neutrino $N_1$. Then, in the context of leptogenesis, the baryon asymmetry is obtained by partial conversion of the leptonic asymmetry via sphaleron interactions. With the correct treatment of flavoured leptogenesis, this conversion reads~\cite{nardi}~:
\begin{eqnarray}
Y_{\cal B} \simeq \frac{12}{37} \sum_{\alpha} Y_{\Delta  \alpha}\ .
\end{eqnarray}
$Y_{\Delta  \alpha}$ is the $B/3-L_{\alpha}$ asymmetry in the lepton flavour  $\alpha$, which is  conserved by sphaleron interactions  and  transmitted to a baryonic asymmetry $Y_{\cal B}$.

\noindent Defining the variable $z=M_1/T$, the BE governing  the abundance of RH neutrinos $Y_{N_1}$,  and the asymmetry $Y_{\Delta _\alpha}$ are \cite{matters}~:
\begin{eqnarray}
\label{n1}
Y_{N_1}^{\prime}(z)&=&- \kappa \left(D(z)+S(z)\right) \left(Y_{N_1}(z)-Y_{N_1}^{eq}(z)\right) \ ,\\
Y_{\Delta  \alpha}^{\prime}(z)&=&-\epsilon_{\alpha} \kappa \left(D(z)+S(z)\right)  \left(Y_{N_1}(z)-Y_{N_1}^{eq}(z)\right)-\kappa_{\alpha} W(z) \sum_{\beta} A'_{\alpha \beta} Y_{\Delta  \beta}(z)\ ,
\label{n2}\end{eqnarray}
where $Y_{N_1}^{eq}$ is the thermal population of the lightest RH neutrino  $N_1$ 
\begin{eqnarray}
Y_{N_1}^{eq}(z)\simeq\frac{45 \, \zeta(3)}{ 2\pi^4 g_{\star}}\frac{3}{4} \, \,z^{2} K_{2}(z)\ , 
\end{eqnarray}
with $g_{\star}=106.75$ in the SM. The CP-asymmetry generated by $N_1$ in the flavour $\alpha$ is given 
by~\cite{issues}:
\begin{eqnarray}
\epsilon_{\alpha}  &  =& \frac{\Gamma_{N1\: \ell_{\alpha}}-\Gamma_{N1 \:\bar{\ell}_{\alpha}}}{\sum_{\alpha}\left(\Gamma_{N1\: \ell_{\alpha}}+\Gamma_{N1 \:\bar{\ell}_{\alpha}}\right) } \nonumber \\ &=&
\frac{1}{(8\pi)}\frac{1}{  [\lambda \lambda^{\dagger}]_{11}}
\sum_{j} {\rm Im}\, \left\{ ( \lambda_{1 \alpha} ) (\lambda
\lambda^{\dagger})_{1j}
 (\lambda^{*}_{j \alpha}) \right\}
g\left(\frac{M_{j}^2}{M_{1}^{2}}\right)\, ,
\end{eqnarray}
where $g$ is the usual loop function~\cite{roulet}. \\
 In eqs. (\ref{n1},\ref{n2}), 
the wash-out parameters have been factorized out, and are defined as follows: 
\begin{eqnarray}
\label{washoutfactor}
\kappa_{\alpha}&\equiv&\frac{\Gamma_{N1\: \ell_{\alpha}}}{H(M_{N_{1}})}=\lambda_{1\,\alpha}\lambda_{1\,\alpha}^{\star}\frac{ v^{2}}{M_{1} m^*} \equiv\frac{\tilde{m_\alpha}}{ m^*}\ ,\\
 \kappa&=&\sum_{\alpha} \kappa_{\alpha}\equiv \frac{\tilde{m}}{ m^*}\, ,
\end{eqnarray}
with $m^*$ the equilibrium neutrino mass, $m^*\simeq 1.08\times10^{-3} \eV$. These parameters exhibit the 
out-of-equilibrium condition on the decay of the right-handed neutrino: the decay process is out-of-equilibrium when the decay rate is slower than the Hubble expansion rate at the temperature $M_{N_{1}}$: $\Gamma \lesssim H(M_{N_{1}})$ and thus $\kappa \lesssim 1$. 
The regime where $\kappa \ll 1 $ is called the weak wash-out regime, in which case  the inverse reactions involving the thermal scatters are rather slow and do not  efficiently wash-out the lepton asymmetry. On the contrary, $\kappa \gg 1$ is the strong wash-out regime, where the inverse reactions strongly wash-out the asymmetry. Depending on the value of this wash-out parameter, analytical approximations of the production and wash-out terms allow us to derive semi-analytical formulae for the baryon asymmetry, as will be see in the next section.
The processes we take into account in eqs. (\ref{n1},\ref{n2}) are  decays and  inverse decays labeled $D(z)$, and $\Delta L=1$ scattering, $S(z)$. Notice that in eq. (\ref{n2}), CP violation in scattering is also taken into account, and thus $S(z)$ further contributes  to the production of the $Y_{\Delta _\alpha}$ asymmetry. In this study we neglect $\Delta L=2$ scatterings (except the real intermediate states already substracted) that are negligible as $M_{N_1}/10^{14}$ GeV $\ll 0.1\times\kappa$ \cite{matters}, as well as well as scatterings involving gauge bosons.
The thermally averaged decay rate is given by:
\begin{eqnarray}
D(z)=z \frac{K_1(z)}{K_2(z)}\ , 
\end{eqnarray}
where $K_n(z)$ are the modified Bessel functions of the $2^\text{nd}$ kind.
The Higgs-mediated scatterings in the $s-$ and $t-$channel contribute to the production of the asymmetry, as well as to the wash-out term. Their effects can be parametrized by two functions $f_1(z)$ and $f_2(z)$: 
\begin{eqnarray}
f_1(z)&=& \frac{S(z)+D(z)}{ D(z)}\simeq \frac{0.1}{ z^2}\left(\frac {15} {8} +z\right) \left[ 1+a_h(z)z^2\log{\left(1+ \frac{0.1}{a_h(z) z}\right)}\right]\, ,
\label{f1}\\
f_2(z)&\simeq& \frac{0.1}{ z^2}\left(\frac {15} {8} +z\right) \left[ \mu(\kappa)+a_h(z)z^2\log{\left(1+ \frac{0.1}{ a_h(z) z}\right)}\right] \, ,\label{f2}
\end{eqnarray}
where $a_h(z) \simeq \log{(\frac{M_1}{M_h(T)})}\simeq \log{(\frac{z}{0.4})}$ and $\mu (\kappa)\simeq 1$ ($2/3$) in the case of weak (strong) wash-out regime. The wash-out term $W(z)$  contains a part from the inverse decay  and a part from scatterings \cite{pedes} and is given by: 
\begin{eqnarray} 
W(z)=W_{ID}(z) f_2(z)\, , \quad
 W_{ID}(z)=\frac{1}{4} z^{3} K_{1}(z) \ .
\end{eqnarray}
The function $f_1(z)$ ($f_2(z)$) parametrizes the effect of the scatterings in the production (wash-out) factor, and  the r.h.s. of eqs.(\ref{f1},\ref{f2})  comes from high-temperature approximations of the reduced-cross sections, when the scattering effects are fully relevant. In the low temperature regime,  scatterings become negligible, so  the functions $f_1$ and $f_2$ tend to unity. 
The matrix $A_{\alpha \beta}$~\footnote{For convenience we use $A'=-A$ in the BE so that the diagonal elements are positive.} depends on which charged lepton interactions are in thermal equilibrium, and parametrizes the conversion  of the leptonic asymmetry into  a $B/3-L_{\alpha}$ asymmetry according to $Y_{\alpha}=\sum_{\beta} A_{\alpha \beta} Y_{\Delta  \beta}$. If the temperature at which leptogenesis occurs, $M_1$,  is  below $10^{9}$ GeV, interactions involving  charged $\mu$ and $\tau$ couplings are fast compared to the Hubble expansion rate, and are therefore in equilibrium. Thus, $\mu$ and $\tau$ flavours have to be treated separately and so the electron flavour is also distinguishable. Then one has \cite{matters}:
\begin{eqnarray}
A=-A'= \left( \begin{array}{ccc}
-151/179 & 20/179 &  20/179 \\ 
 25/358 &  -344/537 &  14/537 \\ 
 25/358 & 14/537 & -344/537
\end{array} \right) \ .
\end{eqnarray}
For $M_1$ between $10^{9}$ GeV and $10^{12}$ GeV, only the charged $\tau$  Yukawa interactions are in equilibrium. The interactions involving the  $e$ and $\mu$ flavours are slower than the expansion rate, so that those flavours are indistinguishable, and the decay of $N_1$ will generate a $Y_{ e+ \mu}$ asymmetry. The  ``flavoured" asymmetries are then reduced to ($Y_{\tau}$-$Y_{e+\mu}$), and  the $B-L\leftrightarrow L$ conversion matrix reads : 
\begin{eqnarray}
A=-A'=\left( \begin{array}{cc}
-417/589 &  120/589 \\ 
 30/589 &  -390/589  \\ 
\end{array} \right)\ .
\end{eqnarray}
In  a recent work~\cite{zeno}, it has been argued that the interaction rates involving the charged Yukawa couplings should be fast compared to the interactions involving the decaying $N_1$ in order to have sufficient  time to project the produced lepton asymmetry onto flavour-space. It has been  derived that  $M_{N_{1}}$ should be below $5\times 10^{11}$ GeV for the tau-Yukawa to be in equilibrium, hence projecting the lepton asymmetry on the ($Y_{\tau}$-$Y_{e+\mu}$) space. This point will be discussed in section 4.\\A formal solution of eq. (\ref{n2}) for the $B/3-L_\alpha$ asymmetry is given by: 
\begin{eqnarray}
\label{solform}
Y_{\Delta  \alpha}(z)&=&-\epsilon_{\alpha} \kappa \int_{z_{in}}^{z} dx D(x)f_1(x) \Delta N_{1}(x) e^{-\kappa_{\alpha} A'_{\alpha \alpha} \int_{x}^{z} dy W(y)}  \\
&-& \kappa_{\alpha} \sum_{\beta \neq \alpha} A'_{\alpha \beta} \int_{z_{in}}^{z} dx W(x) Y_{\Delta  \beta}(x) e^{-\kappa_{\alpha} A'_{\alpha \alpha} \int_{x}^{z} dy W(y) } \, , \nonumber
\end{eqnarray} where $\Delta N_{1}(z)=\left(Y_{N_1}(z)-Y_{N_1}^{eq}(z)\right)$ is the departure from thermal equilibrium.
The first term in eq. (\ref{solform}) had been estimated  for a vanishing initial $N_1$ abundance, $N_{1}(z_{in})=0$, and for an $N_1$ abundance initially at thermal equilibrium \footnote{We will refer to the case of a vanishing initial abundance as a dynamical case, as the population of $N_1$ is created dynamically by thermal processes. The case of a $N_1$ initially at thermal equilibrium will be refered as thermal case, even if it recquires a non-thermal production mechanism.} $N_{1}(z_{in})=N_{1}(z_{in})^{eq}$, in  Refs.~\cite{pedes,matters,BdB}.
The second term, which has been neglected in these previous studies, is responsible for the interdependency of the flavours through the off-diagonal matrix elements $A'_{\alpha \beta}$. 
This term drives a new contribution to the $B/3-L_{\alpha}$ asymmetry and can be relevant. In some cases the latter is in fact the  dominant contribution, as we will later see.\\
We parametrize, as usual, the final asymmetry $Y_{\Delta  \alpha}$ in terms of efficiency factors that contain  all the dependency on the wash-out factors $\kappa$, $\kappa_\alpha$. Those effiencies $\eta_{\alpha}$ are defined by: 
\begin{eqnarray}
\label{eta}
Y_{\Delta  \alpha} &\equiv& - \epsilon_{\alpha}\: \eta_{\alpha}\,Y_{N_1}^{eq}(T \gg M_{N_1})\nonumber \\
&\simeq& -3.9\times10^{-3}\,\epsilon_{\alpha} \left( \eta_{\alpha}^{d}+\eta_{\alpha}^{nd} \right) \ .
\end{eqnarray}
The first term $\eta_{\alpha}^{d}$ has been derived in \cite{matters,BdB}, and its expression will be  presented in the next section. The second term, $\eta_{\alpha}^{nd}$, arises from the non-diagonal conversion of a leptonic flavour, say $L_{\beta}$,  into the  $B/3-L_{\alpha}$ direction and its effect is also studied in the next section. It is clear from eq. (\ref{solform}) that the efficiency $\eta_{\alpha}$ of the process depends on the individual wash-out parameter $\kappa_{\alpha}$, but weighted by the factor $A'_{\alpha \alpha}$ arising from the $B-L \leftrightarrow L$ conversion. We then define $\tilde{\kappa}_{\alpha} \equiv A'_{\alpha \alpha} \kappa_{\alpha}$ (and consequently $\tilde{\kappa} =\sum_{\alpha} \tilde{\kappa}_{\alpha}$ ) as the ``real" wash-out parameter, and thus $\eta_{\alpha}=\eta(\tilde{\kappa}_{\alpha})$. 
\\ The baryon asymmetry reads: 
\begin{eqnarray}
Y_{\cal B}=\frac{12}{37} \sum_{\alpha} Y_{\Delta \alpha}\simeq -1.26\times10^{-3}\,\sum_{\alpha}\epsilon_{\alpha}\: \eta_{\alpha} \ ,\quad \text{with}\quad \eta_{\alpha}=\eta_{\alpha}^{d} + \eta_{\alpha}^{nd}\ .
\end{eqnarray}
The baryon asymmetry is, as the lepton asymmetry, the sum of the diagonal term proportional to $A_{\alpha \alpha}\simeq1$ and of the off-diagonal term proportional  to $A_{\alpha \beta,\  \beta\neq \alpha} \simeq 1/10$. The latter contribution will be negligible for the baryon asymmetry, but will strongly modify the individual lepton asymmetries.
\section{Efficiency factors}
Here we proceed to numerically solve the BE (eqs. (\ref{n1},\ref{n2})) for different configurations of the individual CP asymmetries and wash-out factors, for distinct flavour  ``alignments".
\subsection{Study of the efficiency  $\eta^d$}
In this part, we neglect the off-diagonal part in the last term of eq. (\ref{n2}) and  solve the BE (eqs. (\ref{n1},\ref{n2})). Depending on the initial conditions and on the strenght of the wash-out, several analytical approximations can be derived for $\eta^{d}$ (eq. (\ref{eta})).
\subsubsection{Vanishing initial $N_1$ abundance}
\begin{figure}[htb]
  \centerline{
 \includegraphics[scale=0.25]{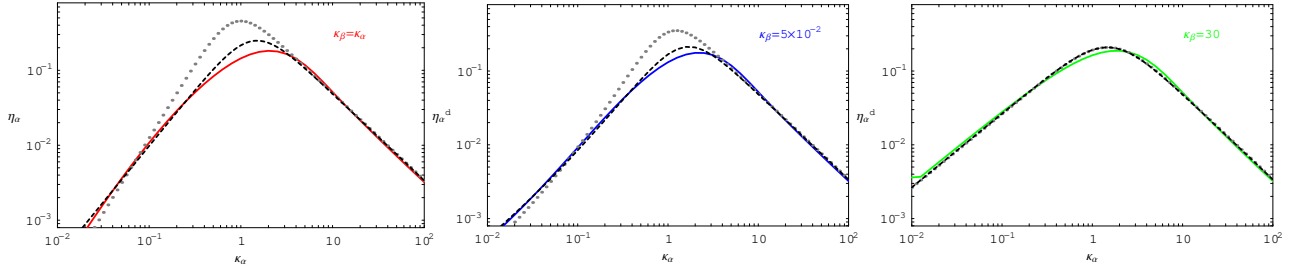}}
\caption{\small Efficiency $\eta^{d}(\tilde{\kappa}_{\alpha})$ in the dynamical case, for specific values of 
 $\kappa_{\beta}$, $\beta\neq\alpha$. Case a) (red curves):~$\kappa_{\beta}~=~\kappa_{\alpha}$~;~case b) (blue curves):~$\kappa_{\beta}=5\times 10^{-2}$;~case c) (green curves):~$\kappa_{\beta}=30$. In each case, the solid lines represent the numerical computation and the dashed ones the results of the analytical approximation. In the left panel, the upper (dotted) curve corresponds to the analytical estimates of the efficiency in the weak wash-out regime eq.~(\ref{faible}).}
 
 \label{figure1}
\end{figure}

In the case where the population of $N_1$ is dynamically generated, i.e. $N_{1}(z_{in})=0$, one can derive the expression for $\eta^{d}$ in different wash-out regimes~\cite{matters}: 
\begin{itemize}
\item all flavours in the strong wash-out case: $\kappa_{\alpha} \gg 1$
\begin{equation}
\eta^{d}(\tilde{\kappa}_{\alpha}) \simeq  3.5 \: \left( \frac{1}{6 \tilde{\kappa}_{\alpha}} \right)^{1.16}\ , \label{fort}
\end{equation}

\item all flavours in the weak wash-out case: $\kappa_{\alpha} \ll 1$\ , 
\begin{equation}
\eta^{d}(\tilde{\kappa}_{\alpha}) \simeq \, \: 1.4 \: \tilde{\kappa}_{\alpha}\ \tilde{\kappa}\: \ .  \label{faible}
\end{equation}
However, within the choosen range of wash-out parameter, we find that the efficiency is better fitted with $\eta^{d}(\tilde{\kappa}_{\alpha}) \simeq  \: 0.4 \: \tilde{\kappa}_{\alpha}\ \sqrt{\tilde{\kappa}} $ , as can be seen in fig.\ref{figure1}.
\item some flavours in strong $\kappa_{\alpha} \gg 1$ and  some others in weak  $\kappa_{\beta} \ll 1$ wash-out regimes. 
In this case the efficiency for the flavour $\alpha$ is given by eq. (\ref{fort}) and the efficiency for the flavour $\beta$ is given by:
\begin{equation}
\eta^{d}(\tilde{\kappa}_{\beta}) \simeq \: 0.3 \:\tilde{\kappa}_{\beta}\ . \label{mid}
\end{equation}
\end{itemize}
The efficiency $\eta^{d}$ is then obtained by simple interpolation between these three generic cases. 
For the sake of illustration, we choose 3 representative cases: 
\begin{itemize}
\item Case a): all the wash-out parameters $\kappa_{\alpha}$ are equal.
\item Case b): some flavours (e.g. $\beta$) are weakly washed-out with $\kappa_{\beta}=5\times  10^{-2}$.
\item Case c): some flavours ($\beta$) are stronlgy washed-out with $\kappa_{\beta}=30$.
\end{itemize}
We checked the validity of those expressions in the considered range of wash-out parameters,   $\kappa_{\alpha}$ between $10^{-2}$ and $10^{2}$.
It is interesting to notice the dependence of $\eta_{\alpha}^{d}$ on  the total wash-out parameter $\kappa$ in eq.~(\ref{faible}).
From this term, a flavour $\alpha$ that is weakly washed-out will be sensitive to the wash-out of the other flavours implying that 
there is a correlation of the flavours. This can be seen in figure \ref{figure1}, where we represent the efficiency of a given flavour ($\alpha$) as a function of the respective   wash-out parameter $\kappa_{\alpha}$, for the three representative cases a), b) and c) discussed above. The democratic scenario, case a), which is similar to the one-flavour approximation, is represented in red and the misaligned cases b) and c) are represented in blue and green, respectively. For comparison, we also represent   the analitycal estimates (dashed lines) of the efficiencies, eqs. (\ref{fort}-\ref{mid}). 
We see that the agreement between the numerical and the analytical results is very good.\\

Firstly, in the case where the right handed neutrino population is created by inverse decays and scatterings, the wash-out factor which parametrizes the strenght of those thermal production, has to be non-negligible in order to produce a sufficient amount of $N_1$. Therefore, the efficiency is maximized in this dynamical case for $\kappa_{\alpha} \simeq 1$ with $\eta_{dyn.}^{max}\simeq 0.2$. \\

Secondly, and this is the main point here, we see that the efficiency of the process when the flavour $\alpha$ is weakly washed-out does indeed depend on the strenght of the wash-out of the other flavours. For example, for $\kappa_{\alpha}\simeq 5\times 10^{-2}$ (and assuming $A'_{\alpha \alpha}\simeq A'_{\beta \beta} \simeq 1$), we roughly have $\eta_{\alpha}^\text{b)}\simeq \eta_{\alpha}^\text{b)} \simeq3\times 10^{-3}$  and $\eta_{\alpha}^\text{c)}\simeq 1.5\times 10^{-2}$.

 The enhancement  $\eta^\text{c)}$ comes from the fact that $\kappa \simeq 10$ but still $\kappa_{\alpha}\lesssim 1$,  and the eq.(\ref{mid}) applies. For cases $a)$ and $b)$, eq.(\ref{faible}) applies, and we have an extra supression from the factor $\kappa \simeq 10^{-1}$. Looking at the strong wash-out regime, $\kappa_{\alpha} > 1 $ , we see that the efficiency of the flavour $\alpha$ does not depend on the wash-out of the other flavours.

\subsubsection{Thermal initial abundance}
In the case where the population of $N_1$ is initially in thermal equilibrium, $N_{1}(z_{in})=N_{1}(z_{in})^{eq}$, the computation of the efficiencies is modified. Following \cite{pedes,BdB}, one has for the efficiency 
factors: 
\begin{eqnarray}
\label{eqth}
\eta^{d}(\tilde{\kappa}_{\alpha})&\simeq&\frac{2}{\tilde{\kappa}_{\alpha}\,z_{B}(\tilde{\kappa}_{\alpha}) f_{1}(z_{B}(\tilde{\kappa}_{\alpha})) }\:\left(1-e^{-\frac{1}{2} \, \left[ {\tilde{\kappa}_{\alpha}\,z_{B}(\tilde{\kappa}_{\alpha})\, f_{1}(z_{B}(\tilde{\kappa}_{\alpha}))}\right]}\right) \ ,
\end{eqnarray}
where 
\begin{eqnarray}
z_{B}(\tilde{\kappa})&\simeq&2+4 \ \tilde{\kappa}^{\, 0.13} \ e^{-2.5\, / \,\tilde{\kappa}}\ .
\end{eqnarray}
\begin{figure}[htb]
  \centerline{
  \includegraphics[scale=0.25]{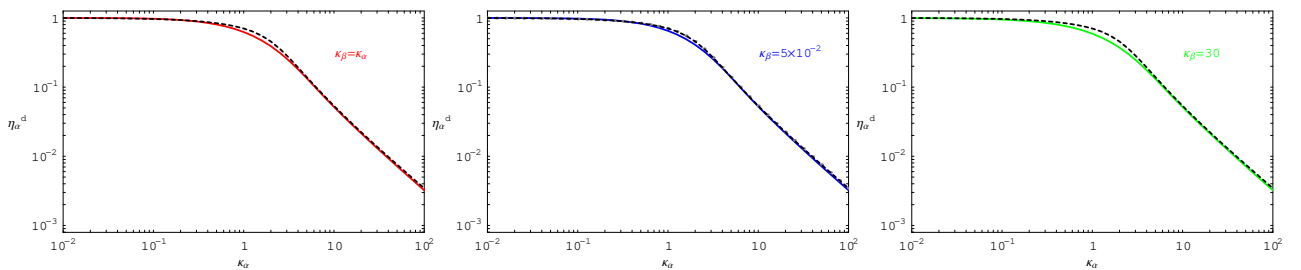}}
\caption{\small Efficiency $\eta^{d}(\tilde{\kappa}_{\alpha})$ in the case of a thermal initial $N_1$ abundance, for specific values of 
 $\kappa_{\beta}$, $\beta\neq\alpha$. Line and color code as in figure \ref{figure1}.}
 \label{grapheTeff}
\end{figure}

The study of the thermal case is shown in  figure \ref{grapheTeff}, where we again consider the 3 cases a), b) and c), with the same colour code as in figure \ref{figure1}. The difference from the case of a vanishing initial $N_1$ is striking in the weak wash-out regime: besides the obvious behaviour of $\eta$ that is maximized for small wash-out, $\eta_{ther.}^{max}\simeq 1$ for $\kappa_{\alpha} \ll 1$, we see that the effect of flavour is negligible, except  for $\kappa_{\alpha}\simeq1$, where a small distinction of the cases a), b) and c) is possible. In the strong wash-out regime, the individual efficiencies do not depend on the wash-out of other flavours, as in the dynamical case. Furthermore, in this strong wash-out regime,  there is no distinction  between the thermal and the dynamical cases, as can be seen in the left panel of figure \ref{figure4} which is included in the discussion of the next section. We also see that the agreement between the analytical result for the efficiency given in eq. (\ref{eqth}) and the numerical result is very good.

\subsection{Study of $\eta^{nd}$}
\begin{figure}[ht]
\begin{center}
\hspace{-1.5cm}
\includegraphics[scale=0.25]{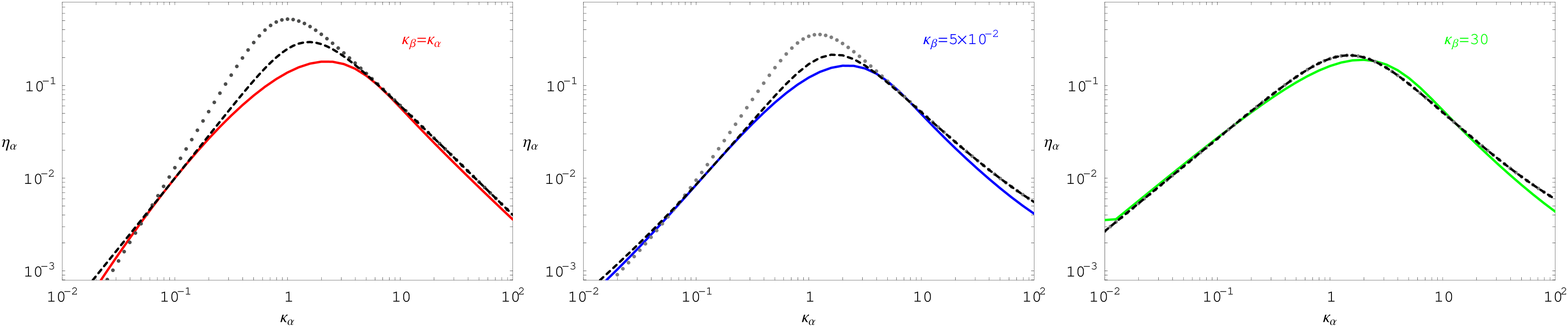} \\ 
\hspace{-2cm}
\includegraphics[scale=0.25]{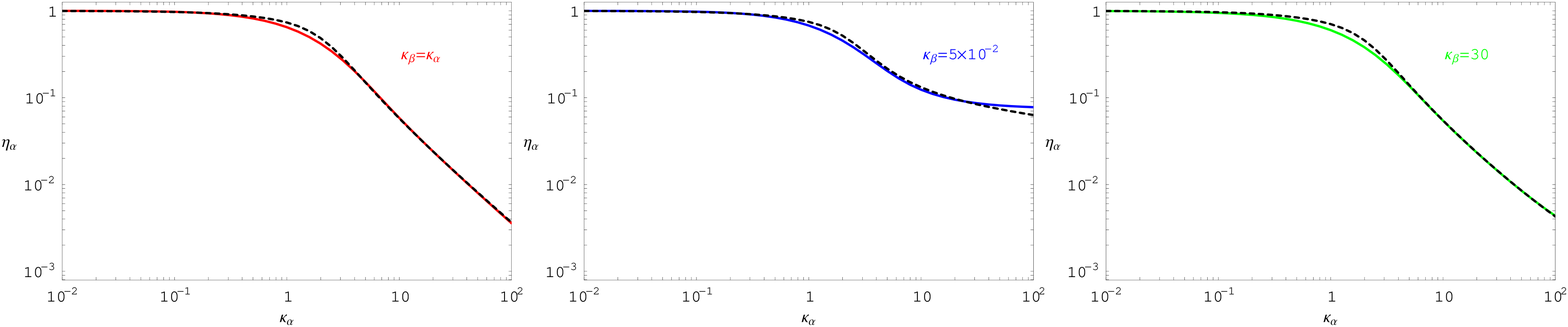}
\end{center}
\vspace{-0.9cm}
\caption{}
{\small 
Influence  of the other flavours on the efficiency $\eta(\tilde{\kappa}_{\alpha})$. Upper panels: case of a vanishing initial abundance. Lower panels: case of a thermal initial abundance. The curves represent different inputs for the wash-out parameters  $\kappa_{\beta}$, $\beta\neq\alpha$: for the  case a)  (red curves) we have a democratic scenario where  $\kappa_{\beta}=\kappa_{\alpha}$, while  for the case b) (blue curves) $\kappa_{\beta}=5\times 10^{-2}$ and for the case c) (green  curves) $\kappa_{\beta}=30$. The solid lines represent the numerical computation and the dashed ones the results of the analytical approximation.
}
\label{grapheMeff}
\end{figure}

\begin{figure}[ht]
\hspace{-1.5cm}
\begin{center}
\includegraphics[scale=0.3]{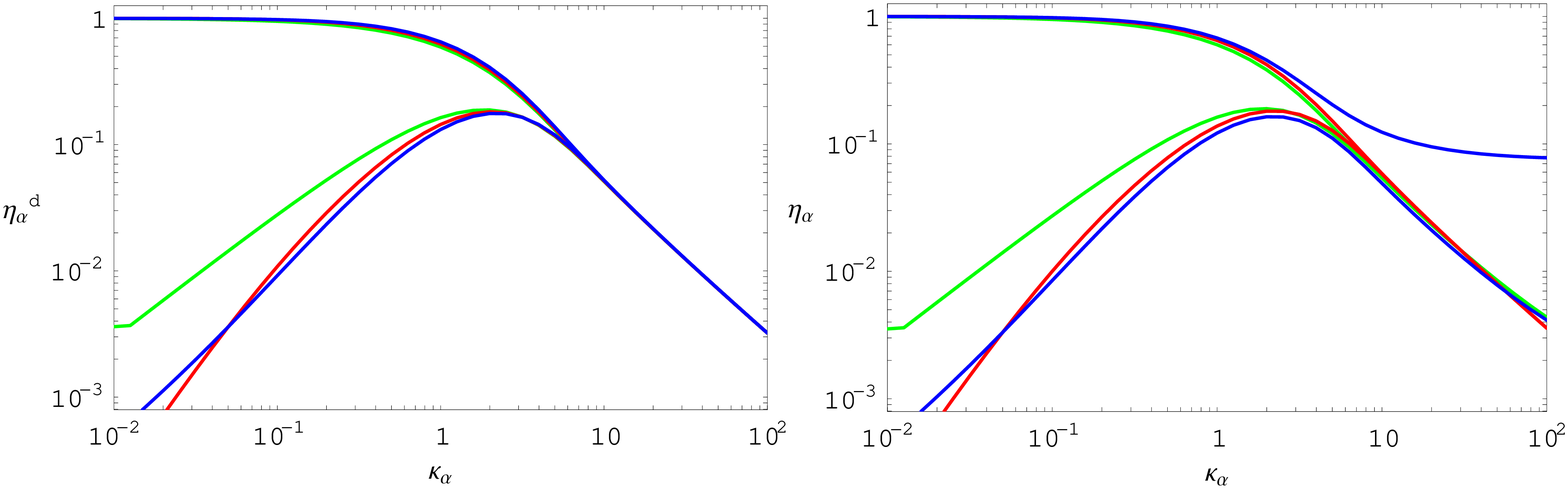}
\end{center}
\vspace{-0.8cm}
\caption{}
{\small 
Influence of the other flavours on the efficiency $\eta(\tilde{\kappa}_{\alpha})$. The left panel represents the diagonal contribution to the efficiency, whereas the right panel represents total, the diagonal and non-diagonal contributions. We represent the case of a zero and thermal initial $N_1$ abundance, for the three cases a), b), c). Line and color code as in figure \ref{figure1}.
}
\label{figure4}
\end{figure}

Now we study the effect of the non-diagonal terms  $A_{\alpha \beta}\ , \alpha \neq \beta$, that exhibit the interplay between different flavours. This interplay  can be seen in the last term  of  eq. (\ref{solform}):
\begin{eqnarray}
\label{depen}
-Y_{\Delta \alpha}^{nd}=3.9\times 10^{-3}\,\epsilon_{\alpha} \eta_{\alpha}^{nd}&=&\kappa_{\alpha} \sum_{\beta \neq \alpha} A'_{\alpha \beta} \int_{z_{in}}^{z} dx\,W(x) Y_{\Delta \beta}(x)\,e^{-\kappa_{\alpha} A'_{\alpha \alpha} \int_{x}^{z} dy W(y) }\ .
\end{eqnarray}
One can find an approximate expression for this term: considering  that  the variations of $Y_{\Delta _\beta}(x)$ are negligible compared to the rest of the integrand, one can approximate it to its final value $Y_{\Delta _\beta}(x)\simeq Y_{\Delta _\beta}(\infty)$, and can thus be factorized out  from the integral. Another approximation is to  consider that  $Y_{\Delta _\beta}$ is mainly generated by the diagonal part, that is we neglect the effects of $\mathcal{O}(A_{nd}^{2})$ that are corrections of the order of  a few percent. We obtain:
\begin{eqnarray}
3.9\times10^{-3}\,\epsilon_{\alpha}\eta_{\alpha}^{nd}&\simeq&\kappa_{\alpha}\, \sum_{\beta \neq \alpha}\, A'_{\alpha \beta}\, Y_{\Delta _\beta}^{d}\:f_{c}( \tilde{\kappa}_{\alpha} )\, ,
\end{eqnarray}
where
 $\tilde{\kappa}_{\alpha}=\kappa_{\alpha} A'_{\alpha \alpha}$ and $f_c(\tilde{\kappa}_{\alpha})$ is given by:
\begin{eqnarray}
\label{offdiag}
f_{c}(\tilde{\kappa}_{\alpha})&=& \int_{z_{in}}^{\infty} dx \ W(x)\  e^{-\tilde{\kappa}_{\alpha} \int_{x}^{\infty} dy W(y) }\\
&\simeq&1.3\,\frac{1}{1+0.8\times\tilde{\kappa}_{\alpha}^{1.17}}\ . 
\end{eqnarray}
The total efficiency of a given flavour is  the sum of the  contribution from the  diagonal part of $A$  which, contains  slight contamination from the other flavours, and from the non-diagonal part, that will be responsible, as we will see, for  a huge modification of the total efficiency,  in some cases becoming dominant compared to the diagonal contribution, 
\begin{eqnarray}
\eta(\tilde{\kappa}_{\alpha})=\eta^{d}(\tilde{\kappa}_{\alpha})+\kappa_{\alpha}\:f_{c}( \tilde{\kappa}_{\alpha} )\, \sum_{\beta \neq \alpha}\, A_{\alpha \beta}\, \frac{\epsilon_{\beta}}{\epsilon_{\alpha}}\:\eta^{d}(\tilde{\kappa_{\beta}})\, .
\end{eqnarray}
The effect of the non-diagonal part on  the  asymmetry $Y_{\Delta_{\alpha}}$  depends on the wash-out  and on the CP asymmetries of the different flavours. For example, if we consider the flavour $\alpha$, the asymmetry produced by the diagonal part proportional to $ \eta^{d}$ depends on the strenght $\tilde{\kappa}_{\alpha}$. 
If $\tilde{\kappa}_{\alpha} \gg 1$, then  $Y_{\Delta_{\alpha}}$ is strongly washed-out and therefore is too small, as can be seen in figure \ref{figure1}: for $\kappa_{\alpha}\gtrsim 100$ we have  $\eta^{d}(\tilde{\kappa}_{\alpha})\lesssim 10^{-3}$. Now consider the non-diagonal part: it depends on the strenght of $\tilde\kappa_{\beta}$. If the wash-out of  the flavour $\beta$ is weak (or even mild), then $\eta^{d}(\tilde\kappa_{\beta})$ will be close to its maximal value, and then one has in this case: 
\begin{eqnarray}
\eta(\tilde{\kappa}_{\alpha})&\propto& \frac{\epsilon_{\beta}}{\epsilon_{\alpha}}\:
\eta^{d}_{\beta \, \text{max}}\times\left(10^{-1}-10^{-2}\right) \nonumber \ ,
\end{eqnarray}
the value of $\eta_\text{max}$ depending on the inital conditions. 
Typically, this effect can, in favourable situations, drive  up the asymmetry $Y_{\Delta_{\alpha}}$  by one or two orders of magnitude as can be seen in the different panels of figure \ref{grapheMeff}. There we  represent the efficiency factor of a given asymmetry $\eta_{\alpha}$ as a function of the wash-out parameter $\kappa_{\alpha}$, for the different alignments of the flavours.
We clearly see that  the off-diagonal terms of $A$ modify the efficiency $\eta_{\alpha}$ in the strong-wash-out case $\kappa_{\alpha}\gg1$. Another consequence of this dependence of $\eta^{nd}_{\alpha}$ on the wash-out of the other flavours is illustrated in figure \ref{figure4}.
We see that when we neglect the off-diagonal terms there is no distinction in the strong wash-out regime between the dynamical and the thermal case. That is, in the strong wash-out regime, the efficiency is independent of the thermal history of the decaying $N_1$. However, when we include the off-diagonal terms, we see that for the case c), the two efficiencies do not coincide anymore in the strong wash-out regime, as each $\eta(\tilde{\kappa}_{\alpha})$ is related to $\eta^{d}(\tilde{\kappa}_{\beta})$, the latter strongly depending on whether the $N_1$ was initially at thermal equilibrium or had a vanishing abundance. This clearly illustrates the effect of flavours in leptogenesis.
\\

\begin{figure}[htb]
  \centerline{
 \includegraphics[scale=0.25]{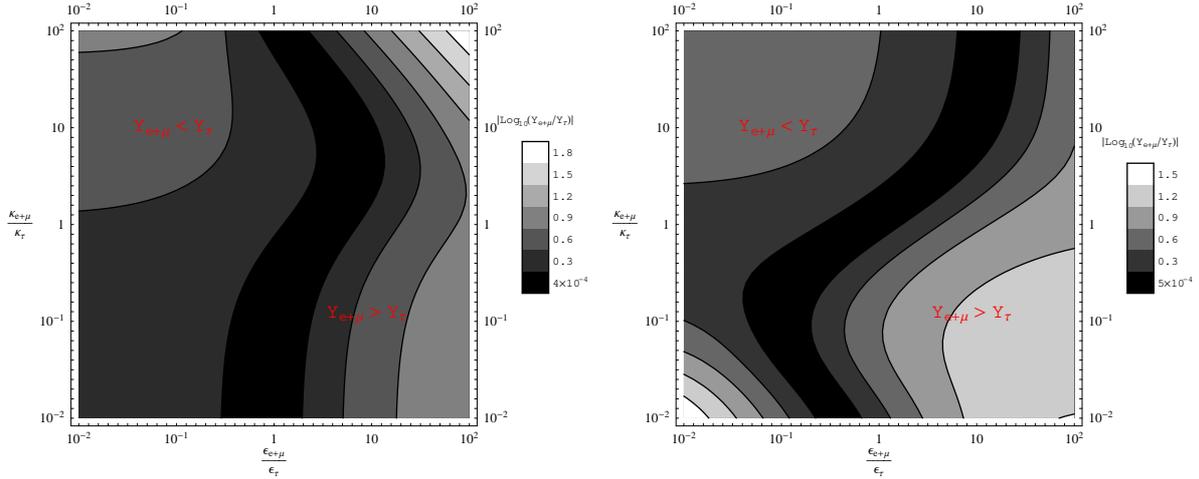}}
\caption{\small Contour plots of the ratio of the individual flavoured asymmetries, plotted as a function of the ratio of wash-out parameter versus the  ratio of CP asymmetries. We place ourselves in a 2 flavour case with $M_{N_1}=5\times10^{9} \GeV$, $\epsilon_{e+\mu}+\epsilon_{\tau}=10^{-6}$ and $\kappa_{e+\mu}+\kappa_{\tau}=10$. The left (right) panel stands for the dynamical (thermal) case. On both graphs, the black area denotes  the equality $Y_{e+\mu}=Y_{\tau}$. Moving to the left  of this countour, $Y_{e+\mu}<Y_{\tau}$ whereas to the right we have $Y_{e+\mu}>Y_{\tau}$.}
 
 \label{figure5}
\end{figure}

The above  discussion is based on a strong approximation ($Y_{\Delta _\beta}(x)\simeq Y_{\Delta _\beta}(\infty)$), which was  made in order to quantify the off-diagonal effect in eq.~(\ref{depen}). 
However this excessively  naive approximation does not describe all wash-out and CP asymmetry configurations. Furthermore, 
 the dynamics of the flavours $\beta\neq \alpha$ is neglected and in particular,  in the case of a strongly washed-out flavour $\beta$, this approximation cannot be used. 
We thus solve the BE,  eqs. (\ref{n1}), (\ref{n2}), including the off-diagonal terms of the $A$ matrix and 
 illustrate in figure \ref{figure5} contours of  the ratio $Y_{e+\mu}/Y_{\tau}$ (absolute value) in logarithmic scales, for the thermal and dynamical cases, as function of the ratio of the flavoured CP asymmetries $\epsilon_{e+\mu}/\epsilon_{\tau}$ and of the flavoured wash-out parameters $\kappa_{e+\mu}/\kappa_{\tau}$. At first sight, we see that $Y_{e+\mu} \geq (\leq) Y_{\tau}$ for $\epsilon_{e+\mu} \geq (\leq) \epsilon_{\tau}$. 
 
 Looking in more detail, we see that the wash-out parameters influence the former statement. For example, in the  thermal case, for $\kappa_{e+\mu}/\kappa_{\tau}\simeq 0.1$ and $\epsilon_{e+\mu}/\epsilon_{\tau}\simeq 0.5$, we have $Y_{e+\mu}\sim 2  Y_{\tau}$.
 
\subsection{The baryon asymmetry}
\begin{figure}[htb]
\centerline{
 \includegraphics[scale=0.3]{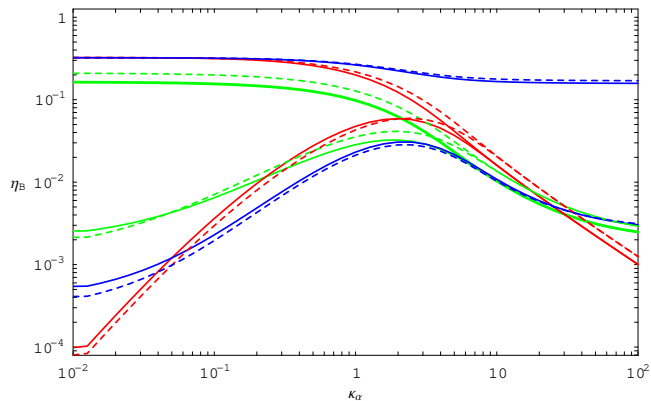}}
\caption{Effects of the off-diagonal terms of the $A$ matrix on the total baryon efficiency $\eta_{\cal B}$.}
\label{yb}
\end{figure}

\begin{figure}[ht]
\hspace{-0.6cm}
\includegraphics[scale=0.4]{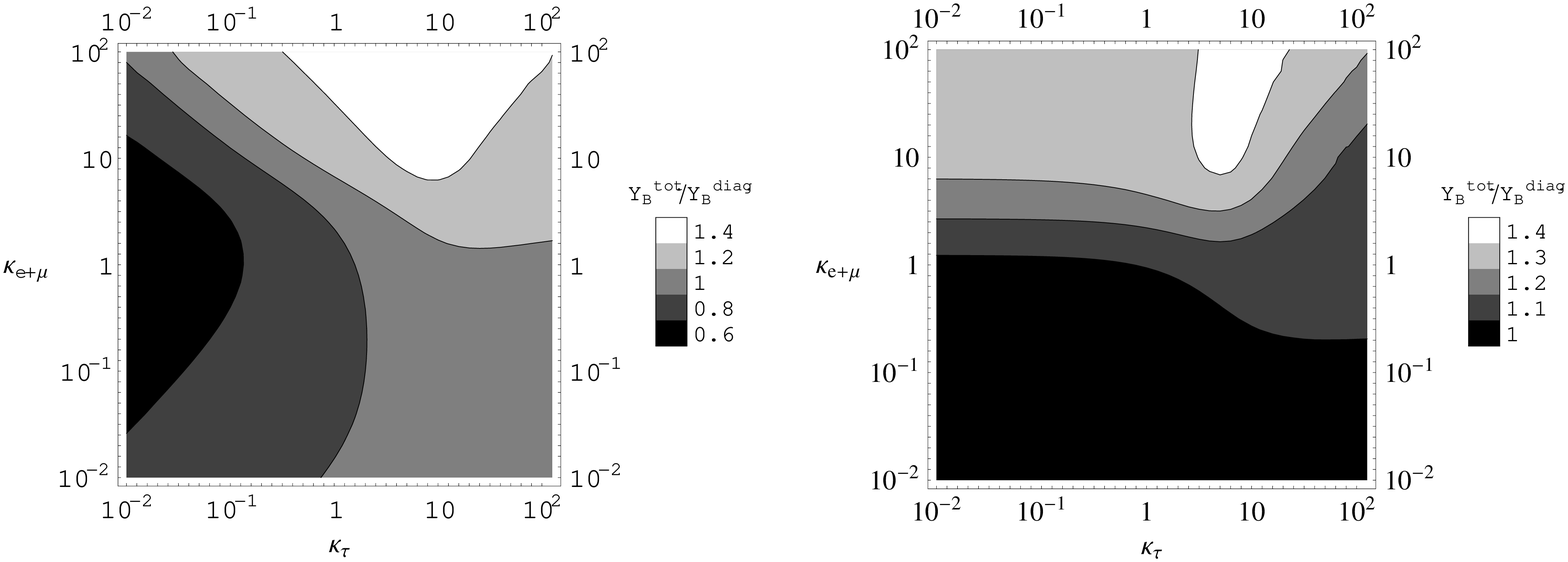}
\vspace{-0.8cm}
\caption{}
{\small 
Effects of the off-diagonal terms on the baryon asymmetry: contour plot of $Y_{B}^{\text {total}}/Y_{B}^{\text {diag.}}$ for fixed CP asymmetries $\epsilon_{e+\mu}=\epsilon_{\tau}$ and varying wash-out parameters. The left (right) panel stands for the case of a vanishing (thermal) initial $N_{1}$ abundance.
}
\label{figureYb}
\end{figure}

Summing-up the contribution from the diagonal and  off-diagonal parts of the $A$ matrix, we finally obtain the baryon asymmetry: 
\begin{eqnarray}
Y_{\cal B} &=&\frac{12}{37} \sum_{\alpha} Y_{\Delta  {\alpha}}\simeq -1.26 \times 10^{-3} \sum_{\alpha} \epsilon_{\alpha}\,\eta_{\alpha}  \, .
\end{eqnarray}
For $\epsilon\equiv \sum_{\alpha} \epsilon_{\alpha}\neq 0$, one can define an efficiency for the baryon asymmetry $\eta_B$ such that:
\begin{eqnarray}
Y_{\cal B} &=&-1.26\times10^{-3} \epsilon \, \eta_{ B} \, . 
\end{eqnarray}
The baryon asymmetry will be the sum of the individual lepton asymmetries. Therefore, the baryon asymmetry, and hence the efficiency $\eta_{B}$, will depend on the alignment of the flavours. We illustrate this point in figure  \ref{yb}, where the efficiency $\eta_{B}$ is plotted  as a function of the wash-out parameter $\kappa_{\alpha}$, for cases a), b) and c). The solid lines represent the diagonal contributions, and the dashed ones represent both diagonal and non-diagonal contributions. As the baryonic efficiency can be defined only for $\epsilon \neq 0$ and is strongly dependent on the flavoured CP violation $\epsilon_{\alpha}$, we set $\epsilon_{\alpha}=8\times10^{-7}$ for all flavours. 
First, we consider only the diagonal contributions. As expected, we see that the efficiency strongly depends on the flavour alignment. The democratic scenario (case a), in red) shows the one-flavour approximation-like behaviour. The flavours  play a full  role in the misaligned case, and especially in  case b) (in blue), where we have $\kappa_{\beta}=5\times10^{-2}$. Then the efficiency for the flavour(s) $\beta$ is (are) close to its (their) maximum, depending on the thermal history of $N_1$. If the flavour $\alpha$ is strongly washed-out, the baryon asymmetry is mainly composed of the flavour(s) $\beta$, and  is thus weakly washed-out, even if the total wash-out is strong. This is the very effect of taking into account the flavours in leptogenesis. We also see that the baryonic efficiency, as well as the leptonic one, depend on the initial abundance of $N_1$ in the strong wash-out regime.\\
Now, let us  consider the dashed lines, which represent the sum of the diagonal and off-diagonal terms of $A$. We see that the off-diagonal terms account for percent corrections. A more detailed analysis of their effect on $Y_{B}$ is shown in fig.\ref{figureYb}, where we represent contours of $Y_{B}^{\text{ total}}/Y_{B}^{\text{diagonal}}$ in the interesting two flavour-case. We choose the individual CP asymmetries to be equal, $\epsilon_{e+\mu}=\epsilon_{\tau}(=10^{-6}$ but the actual value  is not relevant). We see that the off-diagonal terms affect the baryon asymmetry up to $40\%$  in the dynamical case in both ways (increasing and decreasing). On the other hand, the thermal case is only enhanced, up to the same amount. Effects of the off-diagonal terms are particularly important in the strong wash-out regime.
In some particular cases the non-diagonal terms also have a relevant role, namely in the democratic scenario, where $\tilde{\kappa}_{\alpha}$ are equal, and where the total CP asymmetry is zero $\sum_{\alpha} \epsilon_{\alpha}=0$. Then ignoring the $\mathcal{O}(A)$ effect will lead to a vanishing baryonic asymmetry, as $\sum_{\alpha} \epsilon_{\alpha}\: \eta_{\tilde{\kappa}_{\alpha}}^{d}=0$ \cite{lownrjCPV}. On the contrary, taking into account those effects avoids the latter cancellation. For example, in the case of three distinguishable flavours  with the specific   alignment in which $\tilde{\kappa}_{\alpha}=\tilde{\kappa}_{\beta}=\tilde{\kappa}_{\delta}$ and $\epsilon_{\alpha}+\epsilon_{\beta}+\epsilon_{\delta}\equiv\epsilon_{\alpha}(1+b+d)=0 $, one finds for the baryon asymmetry : 
\begin{eqnarray}
Y_{\cal  B}=-1.26\times10^{-3}\,\epsilon_{\alpha}\:\eta^{d}(\tilde{\kappa}_{\alpha})\left(\sum_{\alpha_{1}\neq \alpha}\kappa_{\alpha_{1}}\:f_{c}(\tilde{\kappa}_{\alpha_{1}}) A_{\alpha_{1}\,\alpha}+b \sum_{\beta_{1}\neq \beta}\kappa_{\beta_{1}}\:f_{c}(\tilde{\kappa}_{\beta_{1}}) A_{\beta_{1}\,\beta}+d \sum_{\delta_{1}\neq \delta}\kappa_{\delta_{1}}\:f_{c}(\tilde{\kappa}_{\delta_{1}}) A_{\delta_{1}\,\delta} \right) \ .\nonumber
\end{eqnarray}
In order to maximize the above function, we take $\alpha=\mu$, $\beta=e$ and $\delta=\tau$ for a positive value of  $b$ (in the case of a negative value of $b$, $\alpha=\mu$, $\beta=\tau$ and $\delta=e$). We then get $ Y_{\cal B}\simeq\epsilon_{\mu}\:\eta^{d}(\tilde{\kappa}_{\mu})\:f_{c}(\tilde{\kappa}_{\mu})\:\kappa_{\mu} \times4.5\:\times 10^{-5}\left(1+2 b\:\right) $,
which  is non-zero for  $b\neq -1/2$.
\section{Lower bound on $M_{N_{1}}$ }
When  the flavours are taken into account in leptogenesis, the modification of the asymmetry, combined with the change in  the efficiency factor  may have an impact on the lower bound of  the mass of ${N_{1}}$. From the bound on each individual CP asymmetry~\cite{issues},
\begin{eqnarray}
\epsilon_{\alpha}\lesssim \frac{3 M_{N_{1}} m_\text{max}}{16 \pi v^{2}} \sqrt{\frac{\kappa_{\alpha}}{\kappa}}\ ,
\end{eqnarray}
 one has
\begin{eqnarray}
\vert Y_{\cal B} \vert &\simeq& 1.26\times 10^{-3}\sum_{\alpha}\,\epsilon_{\alpha}\:\eta_{\alpha} \nonumber \\
&\lesssim&1.26\times10^{-3}\,\frac{3\,M_{N_{1}}\,m_\text{max}}{16 \pi v^{2}}\sum_{\alpha} \sqrt{\frac{\kappa_{\alpha}}{\kappa}}\:\eta_{\alpha}\, ,
\end{eqnarray}
from which a lower bound on $M_{N_1}$ is derived,  
\begin{eqnarray}
\label{boundM1}
M_{N_1}&\gtrsim &\frac{16 \pi }{3\times 1.26\times10^{-3}}\frac{v^2}{m_\text{max}} \frac{\vert Y_{\cal B} \vert }{\sum_{\alpha} \sqrt{\frac{\kappa_{\alpha}}{\kappa}}\:\eta_{\alpha}} \\
M_{N_1} &\gtrsim& 7.1\times10^{8} \GeV \left( \frac{m_{\rm{atm}}}{m_\text{max}} \right)\,\left| \frac{Y_{\cal B}}{Y_{\cal B}^{CMB}} \right|   \frac{1}{\sum_{\alpha} \sqrt{\frac{\kappa_{\alpha}}{\kappa}}\:\eta_{\alpha}}\, .
\end{eqnarray}
Since the lower bound on $M_{N_1}$ is inversely proportional to the efficiency $\eta_{\alpha}$, it will therefore depend on the thermal history of the decaying right-handed neutrino. In the case where  $N_1$ are produced by scatterings, the efficiency is maximized for a wash-out $\kappa_{\alpha}\simeq 1$, where $\eta_{\alpha}\simeq 0.2$. In the case where  $N_1$ are non-thermally produced, the efficiency is maximized to $1$ for a very weak wash-out $\kappa_{\alpha} \ll 1$. The lower bound will depend on the alignement of flavours, and in the democratic case one has: 
\begin{eqnarray}
\label{bound}
M_{N_1}\gtrsim \left\lbrace 
\begin{array}{l}
4.1\times10^{8} \GeV\: \ {\rm in\: the\: thermal\: case}\\
2.5\times10^{9} \GeV \:\ {\rm in\: the\: dynamical\: case} \, .
\end{array} \right .
\end{eqnarray}
This bound is close to the one derived in the ``one flavour approximation", where $M_{N_1} \gtrsim 2.1\times10^{8} \GeV$ in the thermal case and $M_{N_1} \gtrsim 1.05\times10^{9} \GeV$ in the dynamical one~\cite{bound}. Besides flavour effects, the difference between the lower bounds of the flavoured and unflavoured cases comes from a different  factor in the  $B-L\leftrightarrow B$ conversion. Indeed, in the one flavour dominance, the Davidson-Ibarra bound reads \cite{DI}: 
\bea
\epsilon \leq \frac{3}{8 \pi} \frac{M_{N_{1}} m_{max}}{v^{2}} \ ,
\eea
and the conversion from sphalerons is $ Y_{B}=28/51 Y_{L} $. Therefore the lower bound on $M_{N_{1}}$ in the unflavoured case is $\sim$  $1/2\times12/37\times 51/28 \sim 0.3$ times the flavoured one.  \\
An implication of this bound resides in the well-known conflict between the reheating temperature and leptogenesis. Indeed, $T_{RH}$ should be above $M_{N_1}$ in order to avoid the erasing of the  lepton asymmetry. In view of the above estimates, $T_{RH} \gsim 4\times 10^{9(8)} \GeV$ in the dynamical (thermal) case. In this case, the inclusion of flavours does not really help. 
In the regime of strong wash-out, $\kappa\gg1$, where the effective neutrino mass is close to the mass inferred from atmospheric oscillations, the situation changes. In the one-flavour approximation, the efficiency $\eta(\kappa)\propto \kappa^{-1}$, therefore $M_{N_1}^{\rm min}\propto \kappa$, and increases with the wash-out, and so does the reheating temperature. In this strong wash-out regime, we roughly estimate  $T_{RH}\geq M_{N_1}/10 \simeq 9\times10^{8} \GeV \: \kappa$, $\kappa\gg1$~\cite{pedes}. Including flavours, one generically has non-alignment of the individual asymmetries, and one can have a strong  total wash-out, with some weakly washed-out flavours. If for example (see fig. \ref{M1m1}), $\kappa\simeq m_{\text{atm}}/m^{*}\simeq 45$ with $\kappa_{\beta}\simeq 40$ but $\kappa_{\alpha}\simeq 5$, one has $M_{N_1}^{\rm min}\simeq 2\times10^{10} \GeV$ for the flavoured case and  $8\times 10^{10} \GeV$ for the unflavoured case. Hence, including flavours, the reheating temperature in this strong wash-out regime is lowered by a factor $\simeq 4$, and one roughly has $T_{RH} \geq 2\times10^{9} \GeV$.
\subsection{Numerical results}
We numerically solve the set of coupled BE (eqs.(\ref{n1}\ref{n2})) and obtain the allowed parameter space. The input parameters of the BE are the Yukawa couplings $\lambda$, which define the wash-out parameters and the CP asymmetries. These  have been built using the Casas-Ibarra parametrization \cite{CI}. In this parametrization the matrix $\lambda$ reads: 
\begin{equation}
\lambda=M_{N}^{1/2}\,R\, m^{1/2} U^{\dagger } v^{-1} ,   \label{yukawa}
\end{equation}
$U$ being  the PMNS matrix parametrized by  three angles and three phases. The solar and atmospheric angles 
are  well measured, whereas the $\theta_{13}$ angle (Chooz angle)  is only  upper-constrained. The three CP violating phases are yet undetermined. The matrix $R$ is a $3\times3$ orthogonal matrix depending on three complex angles. We impose a normal hierarchical spectrum, both for the light and for the heavy neutrino sectors, so that the only free parameters that enter in the  mass matrix  are  the lightest neutrino masses $m_{1}$ and $M_{N_1}$. Therefore, we have 12 independent variables that are not experimentally determined.

\begin{figure}[htb]

\hspace{0.4cm}{
\includegraphics[scale=0.4]{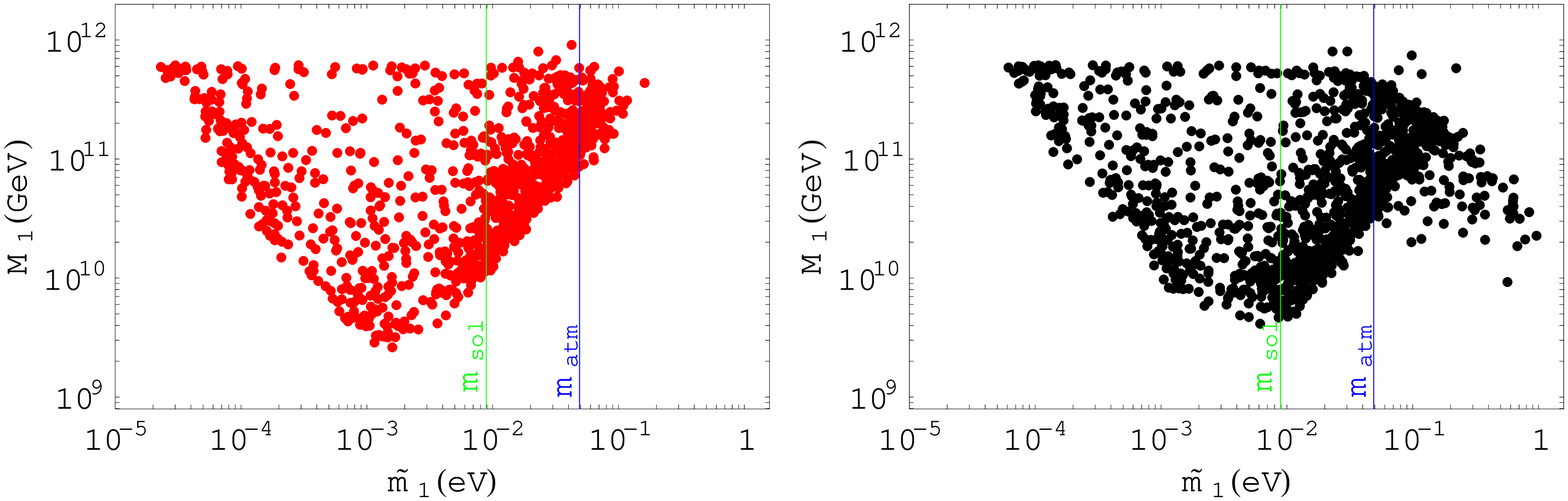}}\\

\hspace{0.4cm}{ \includegraphics[scale=0.4]{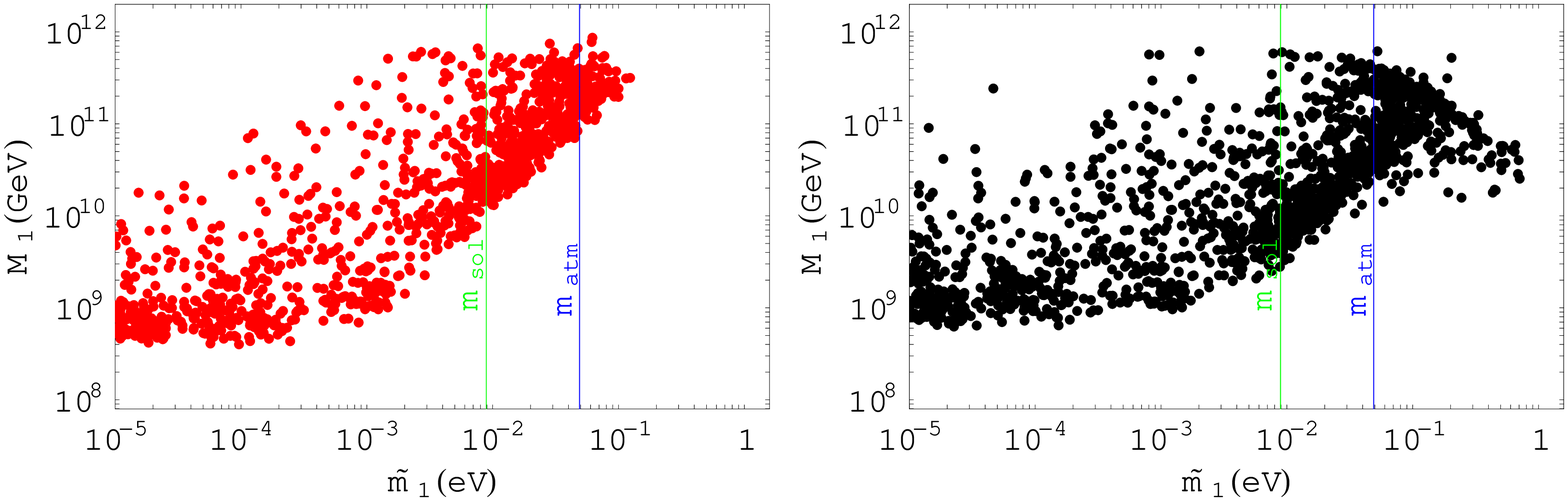}}
\vspace{-0.8cm}
\caption{}
{\small 
Successfull leptogenesis:  bound on $M_1$,  in the case of a zero (thermal) initial $N_1$ abundance for the up (down) panels. The left panels show the allowed ($M_{N_1}$-$\tilde{m}$) parameter space  for the one-flavour approximation, whereas the right panels stand for the case when lepton flavours are taken into account. The vertical lines represent $\sqrt{\Delta m^{2}_\text{atm}}$ (in blue) and $\sqrt{\Delta m^{2}_\text{sol}}$ (in green).
}
\label{M1m1}
\end{figure}

\begin{figure}[ht]
\hspace{0.2cm}
\includegraphics[scale=0.4]{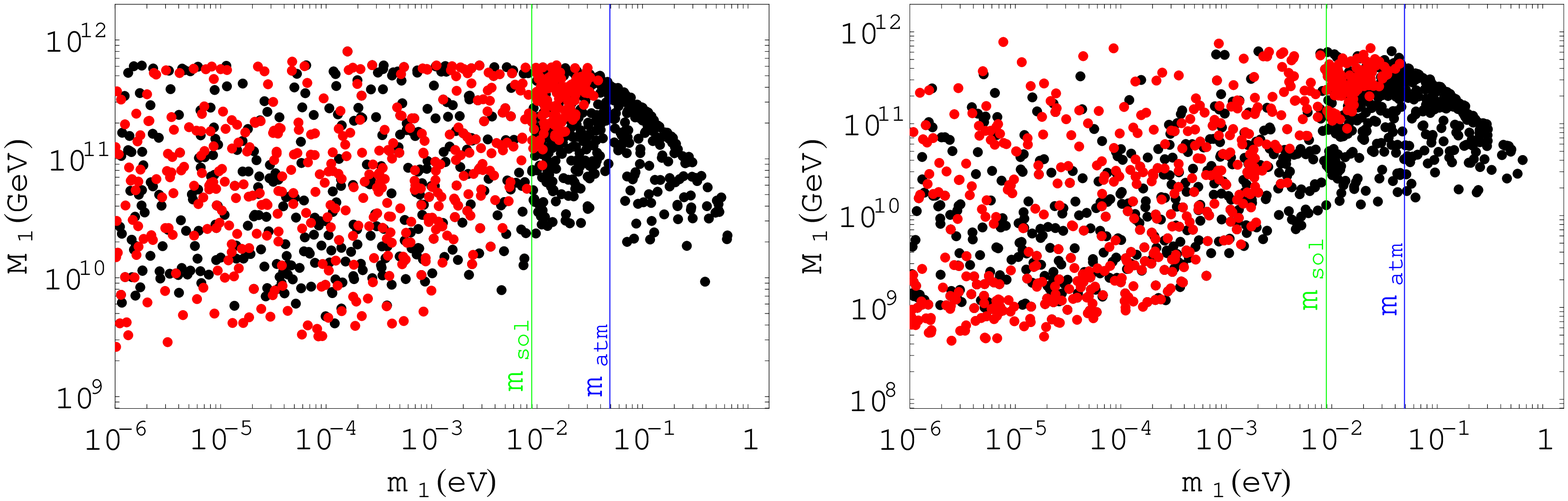}
\vspace{-0.8cm}
\caption{}
{\small 
Successfull leptogenesis:  $M_1$-$m_{1}$ space, in the dynamical case (left panel) and in the thermal case (right panel). The vertical lines represent $\sqrt{\Delta m^{2}_\text{atm}}$ (in blue) and $\sqrt{\Delta m^{2}_\text{sol}}$ (in green).
}
\label{M1m1f}
\end{figure}

In our numerical computation, we scan over the whole parameter space  by  randomly   choosing the free parameters. Moreover, we impose a perturbative limit $\lambda_{33} \lesssim y_{t}$ ($y_{t}$ being  the top Yukawa coupling) that will upper-constraint $M_{N_{1}}$, leading to the upper bound  $M_{N_1}\lesssim 4\times10^{11} \GeV$. 
From \cite{zeno} and  in order for the flavour to be relevant in leptogenesis, an upper bound on  $M_{N_{1}}$ can be derived, namely $ M_{N_1}\lesssim 5.8\times10^{11} \GeV$. Here,  we only  consider  cases were the interactions involving the charged lepton Yukawas are faster than the inverse decay, so that   flavours are fully relevant. Finally, by imposing the  obtained baryon asymmetry to be in the experimental range \footnote{The experimental bounds come from cosmological constraints: the lower  bound corresponds to the  Big-Bang nucleosynthesis (BBN)  constraint on the light species abundance, and the upper-bound from cosmic background radiation (CBR) constraints \cite{BBN}. The WMAP~\cite{wmap} constraints are included therein.}, $5.2\times10^{-10} \lesssim 7.04\times Y_{\cal B} \lesssim 7.2\times10^{-10}$, we  represent in figure \ref{M1m1}  the ($M_{N_{1}}-\tilde{m}$) parameter space allowed by the requirement of a sucessfull leptogenesis.
Many remarks are in order. Firstly, the lower bound numerically derived  confirms the one  given in eq. (\ref{bound}), and we notice that the ``one flavour approximation"  (left  panels in figure \ref{M1m1}) lowers the bound compared to the correct flavour treatment (right panels). Secondly, comparing the dynamical and thermal cases, that is the up and down panels in figure \ref{M1m1}, we see that $\tilde{m}$ can take much smaller values in the case of a thermal initial abundance. As this mass encodes the out-of-equilibrium condition, cf.  eq. (\ref{washoutfactor}), this means that in the thermal case, leptogenesis occurs even for extremely small values of $\tilde m$, 
 while in the dynamical case, this is not possible.
Finally, comparing the left and right panels, one clearly sees the effect of flavour in the ``re-opening" of the  parameter space for higher values of $\tilde{m}$. Indeed, by introducing the lepton flavour asymmetries, we relax the one-flavour approximation that corresponds either to a common behaviour of the individual asymmetries, or to the dominance of one flavour. Now, other configurations are allowed, and the misaligned ones widen the parameter space. The mass $\tilde{m}$ is related to the total wash-out, that is, to the sum of each individual wash-out. In the one-flavour approximation, a high value of $\tilde{m}$ corresponds to a strong wash-out and is disfavored by leptogenesis. On the contrary we observe (cf. figure \ref{yb}) that even if the total wash-out is strong, we can still have flavours that are weakly washed-out, hence dominating the baryon asymmetry and allowing a successfull leptogenesis. Thus, by the inclusion of flavour in leptogenesis no upper-bound on $\tilde{m}$ can be derived.\\ Notice that for $\tilde{m} (m_{1}) \gtrsim \text{atm}$ in figure \ref{M1m1}(\ref{M1m1f}), the points drop below the upper-bound $M_{N_1}\simeq 5\times 10^{11} \GeV$. Indeed, as $m_1\gtrsim \text{atm}$ , $m_{max}\simeq m_{1}\simeq m_{2}\simeq m_{3}$, and the upper-bound on $M_{N_1}$ scales as $1/m_{1}$, c.f eq. (\ref{boundM1}). \\
In the one flavour approximation, a bound on the light neutrino mass was derived in \cite{bound}, and this no longer  holds when flavours are accounted for \cite{issues}.
 However, in \cite{FlavOsc} a bound on the neutrino mass scale of about $2 \eV$ is derived in the flavoured leptogenesis context in the strong wash-out regime and hierarchical wash-out factors $1\ll\kappa_\alpha \ll \kappa_\beta$ and equal CP-asymmetries. 
In this work, we  impose $m_{\nu}$ to be lighter than the cosmological bound $\sum m_{\nu} \lesssim 1 \eV$ and we do not explore configurations leading to higher $m_{\nu}$. 
This can be seen in figure \ref{M1m1f}, which represents the allowed  parameter space ($M_{N_{1}}$-$m_1$ ) in different cases. The  black points are  the result when flavours are included, whereas red ones represent the one-flavour approximation. We clearly see that the  cosmological bound is saturated when flavours are considered, and this does not occur  in the one-flavour approximation. 
In figure \ref{M1m1f}, for $m_1$ above $m_{\text{atm}}$ , the solutions have in general a specific flavour alignement: the flavoured CP-asymmetries are almost equal $\epsilon_{e+\mu}\sim \epsilon_{\tau}$ and  the individual wash-out factors are hierarchical $1\ll\kappa_\alpha\lsim 10 \kappa_\beta$ and the total wash-out is strong. It is well known that such configurations  of wash-out parameters  achieve sucessfull leptogenesis in the flavoured case whereas the unflavoured one fails. For such  specific configurations, effect of the off-diagonal terms of the $A$-matrix on $Y_{B}$  is maximisal (c.f fig \ref{figureYb}) but nevertheless is only a correction without important impact.

\section{Conclusion}
The behaviour of individual lepton asymmetries in the case of vanishing initial $N_1$ abundance has been analysed in~\cite{matters}. In this study we give  semi-analytical results including fine-tuning corrections that depend on flavour alignment. We extend the study to the case of  $N_1$ initially in thermal equilibrium, and confirm that in this case, when  off-diagonal entries of the conversion $B/3-L_{\alpha}\leftrightarrow L$ are neglected, the efficiency factor for a given flavour is independent of the wash-out of other flavours.\\ Independently of the thermal history of the decaying right-handed neutrino, we observed that misalignment of flavours can greatly enhance the baryon asymmetry, when compared to the one-flavour approximation, for an identical  wash-out strenght.We also include off-diagonal entries of the $B/3-L_{\alpha}\leftrightarrow L$ conversion that couple flavours among themselves.
 Even if this inclusion only modifies the baryon asymmetry by  a few percent, thus allowing to safely disregard these terms for $Y_{B}$ computation, we nevertheless stressed  that the individual lepton asymmetries are very sensitive to such interdependencies.
 Finally, we studied the lower bound on the $N_1$ mass and the leptogenesis allowed parameter space. We confirm the lower bound to be  $\sim 4\times 10^{8 (9)}$ GeV for a thermal (vanishing) initial $N_1$ abundance. We have also shown that the parameter space is enlarged, as the flavour (mis)alignment allows for higher values of the wash-out (or equivalently of $\tilde{m}$).
 \section*{Acknowledgements}
 The authors wish to thank S.~Davidson for enlightening discussions and A.M. Teixeira for usefull comments and for reading the manuscript. This project is partly supported by the ANR project  NEUPAC.


\begin{thebibliography}{20}




\bibitem{FY} M.~Fukugita and T.~Yanagida. \newblock 
 Phys. Lett. B {\bf 174} (1986) 45.



\bibitem{leptogen}
 A partial list:~
  M.~A.~Luty,
  Phys.\ Rev.\  D {\bf 45} (1992) 455; A.~Pilaftsis,
  Phys.\ Rev.\ D {\bf 56} (1997) 5431
  [arXiv:hep-ph/9707235]; W.~Buchmuller, P.~Di Bari and M.~Plumacher,
    Nucl.\ Phys.\ B {\bf 643} (2002) 367
[arXiv:hep-ph/0205349];
J.~R.~Ellis, M.~Raidal and T.~Yanagida,
  Phys.\ Lett.\ B {\bf 546} (2002) 228
  [arXiv:hep-ph/0206300];
R.~N.~Mohapatra and S.~Nasri,
Phys.\ Rev.\ D {\bf 71} (2005) 033001
[arXiv:hep-ph/0410369]; Phys.\ Lett.\ B {\bf 552} (2003) 177
  [arXiv:hep-ph/0210271];
 G.~C.~Branco, R.~Gonzalez Felipe, F.~R.~Joaquim, I.~Masina,
   M.~N.~Rebelo and C.~A.~Savoy,
  Phys.\ Rev.\ D {\bf 67} (2003) 073025
  [arXiv:hep-ph/0211001].


  
\bibitem{towards}G.~F.~Giudice, A.~Notari, M.~Raidal, A.~Riotto and A.~Strumia,
Nucl.\ Phys.\ B {\bf 685} (2004) 89 
[arXiv:hep-ph/0310123].

\bibitem{towardsbis}
S.~Antusch and A.~M.~Teixeira,
  JCAP {\bf 0702} (2007) 024
  [arXiv:hep-ph/0611232].



\bibitem{pedes}
W.~Buchmuller, P.~Di Bari and M.~Plumacher,
Annals Phys.\  {\bf 315} (2005) 305
[arXiv:hep-ph/0401240].

\bibitem{DI} S.~Davidson and A.~Ibarra,
  Phys.\ Lett.\ B {\bf 535} (2002) 25.

\bibitem{bound}
 W.~Buchmuller, P.~Di Bari and M.~Plumacher,
    Nucl.\ Phys.\ B {\bf 643} (2002) 367
[arXiv:hep-ph/0205349];
 W.~Buchmuller, P.~Di Bari and M.~Plumacher,
 Phys.\ Lett.\ B {\bf 547} (2002) 128 [arXiv:hep-ph/0209301];
T.~Hambye, Y.~Lin, A.~Notari, M.~Papucci and A.~Strumia,
Nucl.\ Phys.\ B {\bf 695} (2004) 169
[arXiv:hep-ph/0312203].

\bibitem{flav}
R.~Barbieri, P.~Creminelli, A.~Strumia and N.~Tetradis,
Nucl.\ Phys.\ B {\bf 575} (2000) 61 [arXiv:hep-ph/9911315]; 
A.~Pilaftsis and T.~E.~J.~Underwood,
  Nucl.\ Phys.\ B {\bf 692} (2004) 303
  [arXiv:hep-ph/0309342]; 
 T.~Endoh, T.~Morozumi and Z.~h.~Xiong,
  Prog.\ Theor.\ Phys.\  {\bf 111} (2004) 123 
[arXiv:hep-ph/0308276];
 O.~Vives,
  Phys.\ Rev.\ D {\bf 73} (2006) 073006
  [arXiv:hep-ph/0512160].
  
  \bibitem{nardi} 
  E.~Nardi, Y.~Nir, E.~Roulet and J.~Racker,
  JHEP {\bf 0601} (2006) 164
  [arXiv:hep-ph/0601084]; G.~Engelhard, Y.~Grossman, E.~Nardi and Y.~Nir,
  arXiv:hep-ph/0612187;
  E.~Nardi,
  AIP Conf.\ Proc.\  {\bf 917} (2007) 82
  [arXiv:hep-ph/0702033]; Y.~Nir, arXiv:hep-ph/0702199.
  
\bibitem{issues}
A.~Abada, S.~Davidson, F.~X.~Josse-Michaux, M.~Losada and A.~Riotto,
  JCAP {\bf 0604}, (2006) 004 
  [arXiv:hep-ph/0601083].


\bibitem{matters}
A.~Abada, S.~Davidson, A.~Ibarra, F.~X.~Josse-Michaux, M.~Losada and A.~Riotto,
  JHEP {\bf 0609} (2006) 010 
  [arXiv:hep-ph/0605281].

\bibitem{BdB} S.~Blanchet and P.~Di Bari,
  JCAP {\bf 0703} (2007) 018
  [arXiv:hep-ph/0607330];
 S.~Blanchet and P.~Di Bari,
  Nucl.\ Phys.\ Proc.\ Suppl.\  {\bf 168} (2007) 372
  [arXiv:hep-ph/0702089].
  
\bibitem{FlavOsc}
  A.~De Simone and A.~Riotto,
  JCAP {\bf 0702} (2007) 005
  [arXiv:hep-ph/0611357].

\bibitem{zeno}
   S.~Blanchet, P.~Di Bari and G.~G.~Raffelt,
  JCAP {\bf 0703} (2007) 012
  [arXiv:hep-ph/0611337].

\bibitem{roulet} L.~Covi, E.~Roulet and F.~Vissani,
  Phys.\ Lett.\ B {\bf 384} (1996) 169
  [arXiv:hep-ph/9605319].
  
  



  
\bibitem{CI} J.~A.~Casas and A.~Ibarra,
  Nucl.\ Phys.\ B {\bf 618} (2001) 171 
  [arXiv:hep-ph/0103065].


\bibitem{lownrjCPV}
  S.~Pascoli, S.~T.~Petcov and A.~Riotto,
  Nucl.\ Phys.\  B {\bf 774} (2007) 1
  [arXiv:hep-ph/0611338].




\bibitem{BBN} G.~Steigman,
  AIP Conf.\ Proc.\  {\bf 903} (2007) 40
  [arXiv:hep-ph/0611209].

\bibitem{wmap}
  D.~N.~Spergel {\it et al.}  [WMAP Collaboration],
  arXiv:astro-ph/0603449.


\end{thebibliography}
\end{document}